\documentclass[runningheads]{llncs}
\usepackage[T1]{fontenc}
\usepackage{graphicx,verbatim}
\usepackage{multirow}
\usepackage{colortbl}  
\usepackage{xcolor}    
\begin{document}
\title{Augmented Reality-based Guidance with Deformable Registration in Head and Neck Tumor Resection}
\titlerunning{AR-Assisted Deformable Registration for Surgery}
%

\author{Qingyun Yang\inst{1}$^*$ \and 
Fangjie Li\inst{1}$^*$ \and
Jiayi Xu\inst{1} \and
Zixuan Liu\inst{1} \and
Sindhura Sridhar\inst{2} \and
Whitney Jin\inst{2} \and
Jennifer Du\inst{2} \and
Jon Heiselman\inst{1} \and
Michael Miga\inst{1} \and
Michael Topf\inst{1,2} \and
Jie Ying Wu\inst{1}$^\dagger$ 
}

\def\thefootnote{*}\footnotetext{These authors contributed equally to this work}
\authorrunning{Q. Yang \& F. Li et al.}
%

\institute{Vanderbilt University, Nashville TN 37235, USA
\and
Vanderbilt University Medical Center, Nashville TN 37232, USA\\
$^\dagger$  corresponding author \email{JieYing.Wu@vanderbilt.edu}}

\maketitle              
%

\begin{abstract}
Head and neck squamous cell carcinoma has one of the highest rates of recurrence. Recurrence rates can be reduced by accurate localization of positive margins. While frozen section analysis of resected specimens provides accurate intraoperative margin assessment, complex 3D anatomy and significant shrinkage of resected specimens complicate margin relocation from the specimen back to the post-resection cavity. We propose a novel deformable registration framework that uses both the pre-resection external surface and the post-resection cavity of the specimen to incorporate thickness information. In tongue specimens, the proposed framework improved the target registration error (TRE) by up to 33\% as compared to using the post-resection cavity alone. 
We found distinct deformation behaviors in skin, buccal, and tongue specimens, highlighting the need for tailored deformation strategies. Notably, tongue specimens hold the highest clinical need for improvement among head and neck specimens. To further aid intraoperative visualization, we also integrated this framework into an augmented reality-based guidance system. This system can automatically overlay the deformed 3D specimen mesh with positive margin annotation onto the post-resection cavity. The integrated system improved a surgeon and a trainee’s average relocation error from 9.8 mm to 4.8 mm in a pilot study. Our implementation code for AR guidance and generating the target point cloud is available at https://github.com/vu-maple-lab/Head-and-Neck-Tumor-Resection-Guidance.

\keywords{Augmented Reality \and Deformable Registration \and Head and Neck Cancer \and Positive Margin Relocation}

\end{abstract}
\section{Introduction}
Approximately 890,000 new cases of head and neck squamous cell carcinoma (HNSCC) are diagnosed worldwide each year~\cite{Barsouk2023EpidemiologyCarcinoma}. HNSCC has one of the highest positive margin rates~\cite{Orosco2018PositiveCancers}. Positive margin findings lead to increased local recurrence, reduced survival, and higher treatment costs. Intraoperative frozen section analysis is used to detect positive or close margins. If positive or close margins are detected on the specimen, the surgeon performs further resection on the patient. Thus, precise relocation of positive or close margins from histopathological analysis onto the patient is critical for effective re-resection. However, the complex 3D anatomy of the head and neck region and the variable shrinkage of the specimen complicate accurate relocation. In addition, histopathological findings are typically communicated only verbally to the surgeon, which provides limited spatial information~\cite{nayanar2019frozen}. Only 29\% of re-resections following an initial positive margin contain additional cancer~\cite{prasad2024often}.

To address this unmet need for better guidance in margin relocation, our group has previously introduced an augmented reality (AR) system for intraoperative visual guidance. The AR system enables surgeons to overlay 3D specimens, obtained from intraoperative structured light scans, onto the post-resection cavity~\cite{Tong2024DevelopmentResection}. However, a major challenge of this protocol is that surgical specimens can deform anywhere from 11.3 to 47.6\% during resection~\cite{Umstattd2017ShrinkageFixation,Johnson1997QUANTIFICATIONCAVITY,Mistry2005Post-resectionSignificance,El-Fol2015SignificanceCarcinoma}. The deformation could impact the surgeon's ability to relocate a positive margin. In prior work~\cite{Yang2025Nonrigid}, we used an RGBD camera to acquire a point cloud of the post-resection cavity (resection cavity after tissue removal) as deformation targets. However, a single-surface point cloud does not provide an adequate registration constraint, especially for specimens with complex, non-planar shapes and varied thickness, such as tongue specimens. In this study, we focus on three important tissue types in the head and neck area: skin, buccal, and tongue. Skin and buccal specimens are thin, with pre-resection external surfaces nearly parallel to the underlying post-resection cavity. In contrast, the tongue is structurally thick and exhibits a more complex 3D anatomical shape. Buccal and tongue specimens undergo the most significant mucosal shrinkage post-resection~\cite{necker2023virtual}, with tongue posing the greatest clinical challenge in margin relocalization. 

In this work, we make the following contributions:
(1) A novel deformable registration framework for head and neck tissue, incorporating the pre-resection external surface of the specimen as an additional registration constraint alongside the post-resection cavity. Evaluation results demonstrate that the proposed registration framework offers greater adaptability for thicker specimens.
(2) To the best of our knowledge, we present the first system that combines deformable registration with an AR head-mounted display (HMD) guidance platform to enable intraoperative margin relocalization in head and neck surgery.
(3) Evaluate the target registration error (TRE) between the resected histopathological specimen and the intraoperative post-resection cavity on nine specimens. We also conduct a pilot study to assess surgeons’ ability to relocate a target margin.

\section{Methods}
\begin{figure}[tb]
    \centering
    \includegraphics[width=1\linewidth]{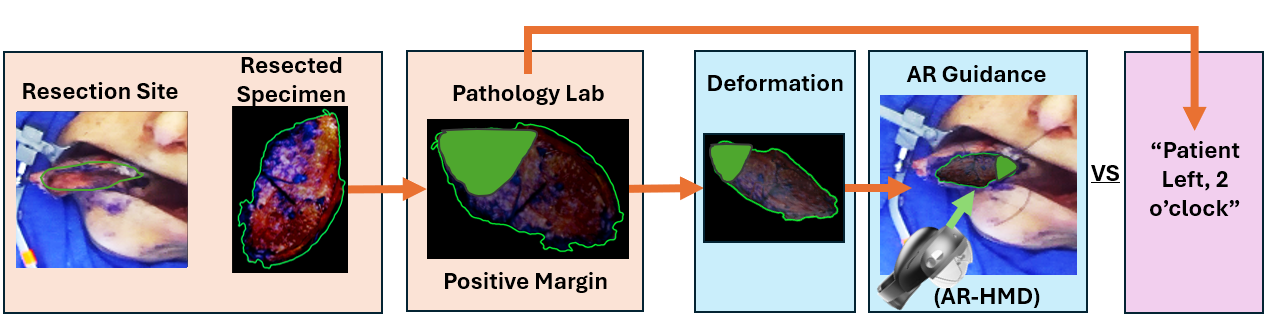}
    \caption{The standard of care and proposed tumor resection workflows. In the standard of care, the surgeon relies purely on verbal instructions to locate the positive margin, leading to imprecise re-resection. AR guidance with deformable registration provides additional visual guidance.}
    \label{fig:general_workflow}
\end{figure}

The proposed deformable registration and AR guidance framework is outlined in Fig.~\ref{fig:general_workflow}, and contrasted with the current standard of care: verbal guidance. 

\begin{figure}[tb]
    \centering
    \includegraphics[width=1\linewidth]{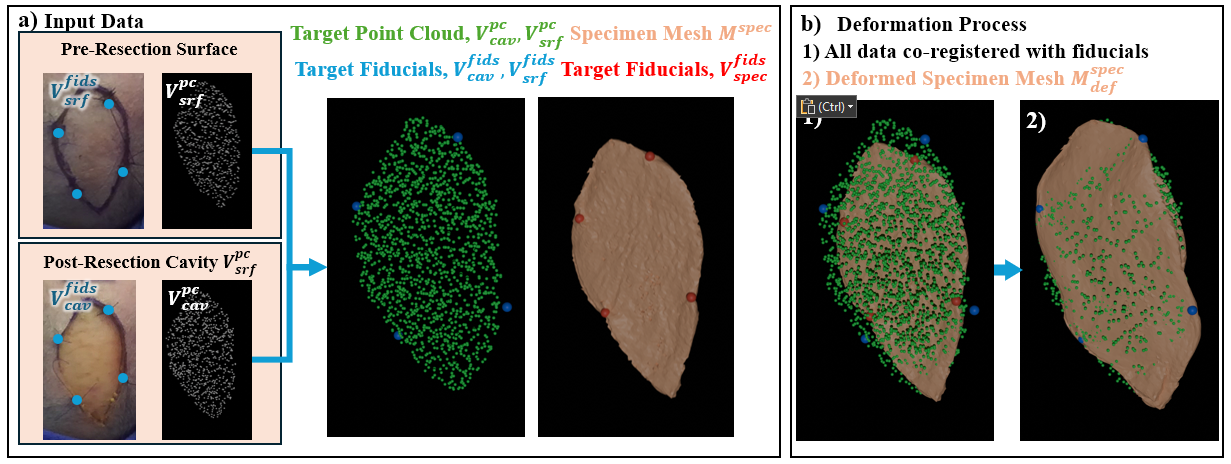}
    \caption{a) A range of input data for the deformation registration task. b) The expected output of the deformation task.}
    \label{fig:deformation_formulation}
\end{figure}

\textbf{Problem Formulation}
Given a 3D mesh of the resected tumor specimen and point clouds of the pre-resection external surface and post-resection cavity, we seek to register the specimen to the pre-resection external surface and the post-resection cavity. Fig.~\ref{fig:deformation_formulation} shows the deformable registration pipeline. The pipeline takes as input: 
\begin{enumerate}
    \item Textured point clouds of the pre-resection external surface of the surgical site, and post-resection cavity, $V^{pc}_{srf}$ and $V^{pc}_{cav}$
    \item Dense 3D mesh of the resected specimen, $M^{spec}$ 
    \item Four corresponding fiducial markers present on the pre-resection external surface $V^{fids}_{srf}$, post-resection cavity $V^{fids}_{cav}$, and the specimen scan $V^{fids}_{spec}$ 
\end{enumerate}
We aim to produce a deformed mesh of the specimen, $M^{spec}_{def}$, registered to the post-resection cavity. In our method, we rigidly align $V^{pc}_{srf}$ and $V^{pc}_{cav}$ using fiducials, $V^{fids}_{srf}$ and $V^{fids}_{cav}$ to produce the target point cloud, $V^{pc}_{targ}$. We use $V^{pc}_{targ}$ as the deformation target, instead of only $V^{pc}_{cav}$ as in~\cite{Yang2025Nonrigid}. We hypothesize that the additional constraint would better guide the deformation process.

\textbf{Deformable Registration Algorithm} We use a Kelvinlet-based deformable registration~\cite{RingelMorganandHeiselman2023RegularizedBreast}, which computes closed-form solutions for elastic deformations based on a Kelvinlet-based physical model. To avoid unrealistic deformations, we incorporate strain energy based regularization into the optimization goal. An isotropic scaling factor is also incorporated, following the approach from \cite{heiselman2021improving}, to correct for uniform tissue shrinkage. For the Kelvinlet model, we used $k=45$ control points, a radial scale parameter of 0.01 m, a strain energy regularization weight of $10^{-11} Pa^{-2}$, a Poisson's ratio of 0.45 and Young's modulus of 2100 Pa.



\begin{figure}[tb]
    \centering
    \includegraphics[width=0.90\linewidth]{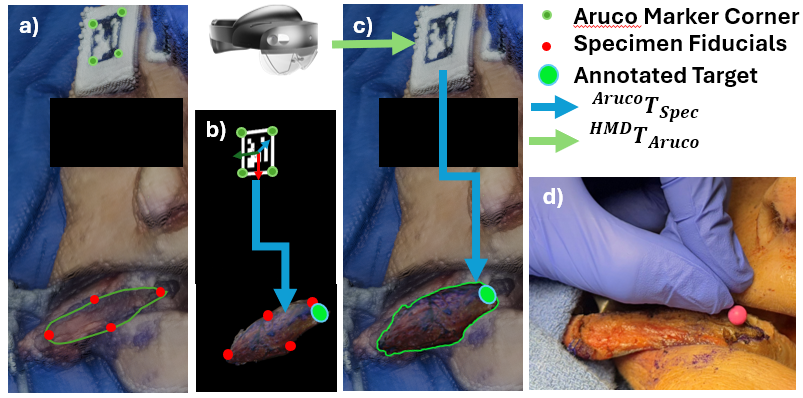}
    \caption{General Workflow of the AR system. a) From post-resection cavity point cloud extracted from the RGB-D camera, we obtain the pose of the post-resection cavity and the ArUco marker. b) The transformation between the ArUco marker and the post-resection cavity. c) The surgeon wearing the HMD can visualize the annotated specimen mesh, automatically overlaid on the post-resection cavity. d) The surgeon places a surgical pin on the target to evaluate end-to-end accuracy.}
    \label{fig:AR_workflow}
\end{figure}

\textbf{Automatic Registration}
The process to overlay the specimen mesh onto the post-resection cavity is detailed in Fig.~\ref{fig:AR_workflow}. An ArUco marker~\cite{Garrido-Jurado2014AutomaticOcclusion} is affixed to the surface of the cadaver head prior to resection. Following resection, the pose of both the ArUco marker and the post-resection cavity are obtained through the post-resection point cloud $V^{pc}_{cav}$. Using these poses, the deformed mesh is registered to the ArUco marker and uploaded to the HMD. The HMD continuously tracks the ArUco marker on the cadaver surface, allowing the virtual, annotated mesh to be overlaid onto the post-resection cavity, enabling visual guidance. 

\textbf{System Requirements}
We used a regular laptop with an NVIDIA GeForce RTX 4070 GPU for the deformable registration framework. Including the time required to scan both the specimen and the cavity, the entire workflow from data acquisition to deformable registration, AR-based automatic registration, and deployment can be completed within 30 minutes. After deformable registration, we use a HoloLens 2 (Microsoft, Seattle, WA) to superimpose a rendered overlay of the deformed specimen onto the post-resection cavity. The AR interface is developed with Unity 3D 2019 and the Microsoft Mixed Reality Toolkit (MRTK).

\section{Experiment Setup and Data Collection}

\textbf{Data Acquisition for Deformable Registration} We collected three skin, three buccal, and three tongue specimens to evaluate the proposed method on a variety of tissue types from three fresh-frozen human cadaver heads. A head and neck cancer surgeon, who performs approximately 50 locally advanced head and neck surgical resections annually, resected specimens similar to those in clinical cases. Before resections, he used an ink pen to mark the surgical plan. He added four pairs of sutured stitches at the corresponding locations on the boundaries of both the specimen and the post-resection cavity. These paired stitches served as fiducial markers ($V^{fids}_{srf}$, $V^{fids}_{cav}$, and $V^{fids}_{spec}$) for deformable registration.

We used a structured light 3D scanner (EinScan SP, Shining 3D, China) to generate a 3D, textured mesh of the resected specimen~\cite{Perez2023ExPathology}. We used an RGB-D camera (ZED 2i, Stereolabs Inc., USA) to capture sparse colored point clouds of the pre-resection external surface and post-resection cavity. We attached the camera to a mechanical arm, allowing the camera to operate overhead, without interfering with the surgeons~\cite{ringel2024image,Yang2025Nonrigid}. We manually segmented the pre-resection external surface and post-resection cavity from the RGB-D images and labeled the fiducial markers on the mesh and the point clouds (Fig.~\ref{fig:deformation_formulation}a).

\textbf{Deformable Registration Evaluation} We evaluated the method with leave-one-out cross validation. For each fold, we excluded one fiducial marker from being used for the rigid initialization of the registration. We then evaluated the TRE, the distance between the estimated fiducial marker location after registration and the location of the corresponding fiducial marker on the post-resection cavity. As baseline, we also compared to rigid and similarity registration (rigid registration
with scaling). Paired t-tests between methods were performed.

\textbf{End-to-End Evaluation} Tumor recurrence rate is inversely correlated with re-resection performance \cite{Orosco2018PositiveCancers}, which is dependent on surgical margin relocation accuracy \cite{Miller2024HowNeck}. To evaluate the clinical potential of our system, a surgeon and a trainee relocated two specimens on the post-resection cavity. We randomly chose a pair of fiducial markers from two specimens as a target proxy for the margin to be relocated. We first measured the target fiducial marker ground truth positions on its post-resection cavity based on landmarks. Then, we removed the marker. For each specimen, the surgeon and trainee first performed the task with only verbal guidance (e.g., “patient left inferior 6 o’clock”). Then, they repeated the task with AR guidance. We measured the Euclidean distance between their identified position and the ground truth of the target fiducial marker.

\section{Results}
\textbf{Deformable registration framework} 

\begin{figure}[tb]
    \centering
    \includegraphics[width=1\linewidth]{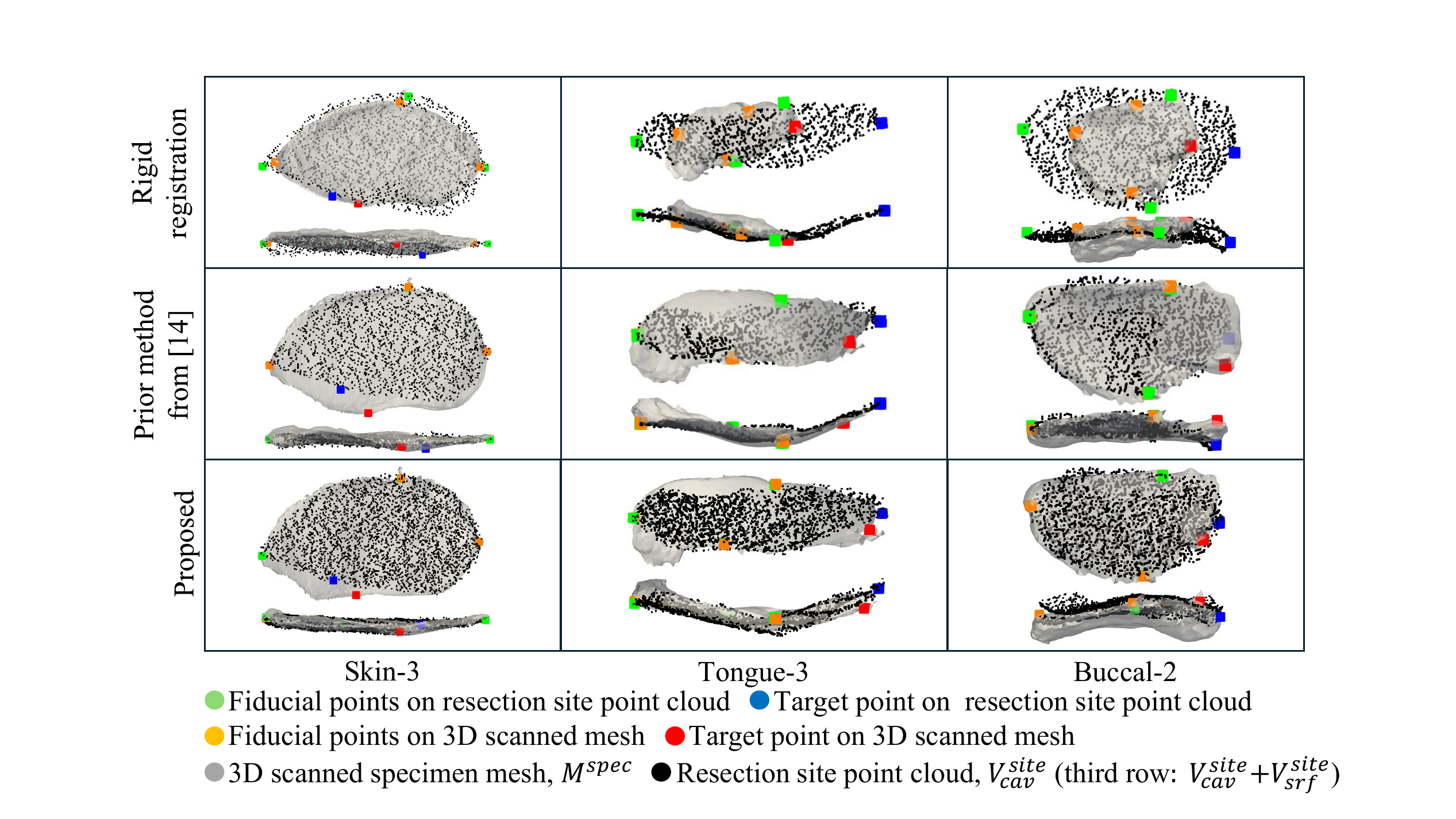}
    \caption{The 3D specimen with rigid registration to post-resection cavity (first row), deformable registration with a post-resection cavity guidance (second row), and the proposed deformable registration with additional pre-resection external surface guidance (third row), respectively.}
    \label{fig:specimen_results}
\end{figure}

Fig.~\ref{fig:specimen_results} presents the deformation results for all three specimen types. Deformable registration showed improved alignment across all specimens, but the degree of improvement varied across specimen types. For skin, deformable registration led to only modest refinement, likely due to its initially small rigid registration error. In contrast, tongue and buccal specimens exhibited substantial improvement at the target points, reflecting the greater deformation present.

\begin{figure}[tb]
    \centering
    \includegraphics[width=0.7\linewidth]{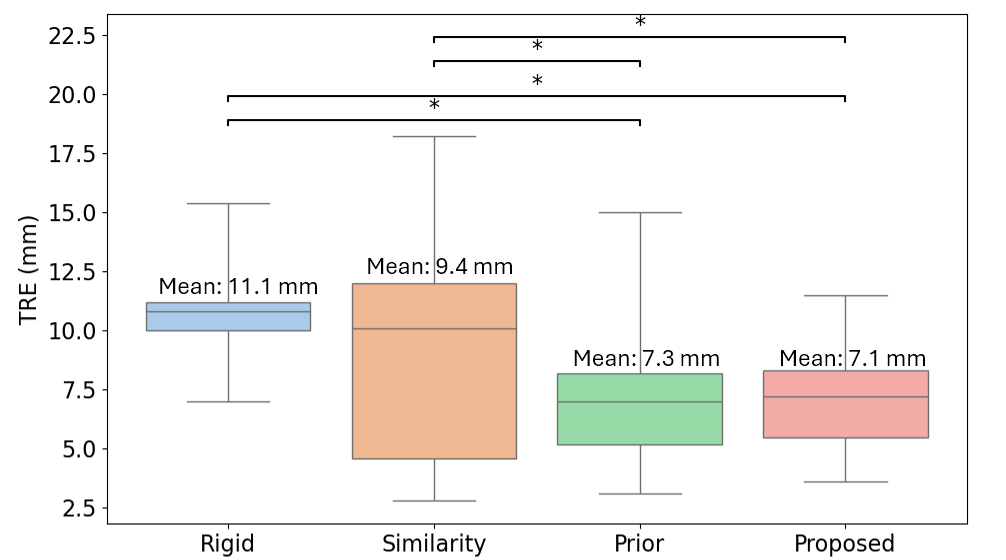}
    \caption{The overall TRE (mm) of four different registration methods. The proposed method significantly outperforms the rigid and similarity registrations, which was not possible with the previous method. *: $p<0.05$.}
    \label{fig:result_box_plot}
\end{figure}

Fig.~\ref{fig:result_box_plot} shows that both deformable registration methods achieved significantly lower average TRE than rigid and similarity registration ($p \leq 0.05$) across all specimens. In two cases, similarity registration yielded a lower TRE (Table \ref{tab:result_tab}). However, similarity registration exhibited high variability and was not significantly more accurate than rigid registration (p > 0.05) in buccal and tongue specimens. There was no significant difference between the two deformable registration methods. However, the maximum TRE and variance are lower for our proposed method.

Across all three tongue cases, the proposed approach consistently achieved at least 20\% lower TRE than the prior deformable method. Conversely, for buccal specimens, the prior deformable method consistently achieved a lower TRE across all three cases. Notably, the proposed method did not improve performance compared to rigid registration in this category. For skin specimens, the advantage of the proposed method was less consistent, outperforming the prior method in 2 out of 3 cases.

\begin{table}
\centering
\caption{TRE (mm) of All Specimens, Grouped by Anatomical Source.}\label{tab:result_tab}
\begin{tabular}{|l|l|l|l|l|}
\hline
\rowcolor{gray!20}Specimen& Rigid & Similarity & Prior method from~\cite{Yang2025Nonrigid} & Proposed  \\
\hline
Skin-1& $10.0\pm1.1$ & $4.6\pm1.4$ & $7.0\pm0.9$ & $\textbf{3.6}\pm1.5$\\
\rowcolor{gray!15}Skin-2& $10.8\pm3.8$ & $10.1\pm1.7$ & $\textbf{5.2}\pm$3.2 & $7.2\pm5.0$\\
Skin-3& $11.2\pm1.5$ & $10.9\pm2.3$ & $8.4\pm4.2$ & $\textbf{8.3}\pm1.9$\\
\rowcolor{gray!15} Buccal-1& $10.5\pm6.0$ & $13.6\pm3.9$ & $\textbf{8.2}\pm4.4$ & $8.7\pm3.7$\\
Buccal-2& $7.0\pm3.1$ & $\textbf{2.8}\pm0.9$ & $4.9\pm1.0$ & $7.2\pm1.5$\\
\rowcolor{gray!15}Buccal-3& $11.0\pm1.7$ & $8.7\pm3.1$ & $\textbf{3.1}\pm1.2$ & $7.2\pm3.3$\\
Tongue-1& $15.4\pm2.4$ & $18.2\pm2.6$ & $15.0\pm5.9$ & $\textbf{11.5}\pm3.1$\\
\rowcolor{gray!15} Tongue-2& $9.2\pm4.6$ & $\textbf{3.7}\pm1.9$ & $5.4\pm2.3$ & $4.3\pm1.2$\\
Tongue-3& $14.9\pm7.4$ & $12.0\pm1.0$ & $8.2\pm3.2$ & $\textbf{5.5}\pm5.1$\\
\hline
\end{tabular}
\end{table}

\textbf{User Study}
Table \ref{tab:ar_result_tab} shows that in all user study cases, the relocation accuracy with our guidance system was better or no worse than with verbal guidance. On average, relocation accuracy improved by 46.3\%. 

\begin{table}
\centering
\caption{TRE (mm) Between Surgeon-Relocated Target Points and Ground Truth Positions}\label{tab:ar_result_tab}
\begin{tabular}{|l|l|l|l|l|}
 \hline
\rowcolor{gray!20}Guidance Type& Skin-1, User 1 & Skin-1, User 2 & Tongue-2, User 1 & Tongue-2, User 2  \\
\hline
Verbal& 10 & 7.4 & 13 & \textbf{8.7}\\
\rowcolor{gray!15}AR& $\textbf{6.1}$ & \textbf{3.4} & \textbf{1.0} & \textbf{8.7}\\
\hline
\end{tabular}
\end{table}

\section{Discussion}

\textbf{Deformable Registration Framework}

The proposed method, which incorporates the pre-resection external surface to guide deformable registration, outperformed the previous method for all tongue specimens, which are of greatest clinical relevance. In skin specimens, it achieved comparable performance to the prior method, while for buccal specimens, the previous deformable registration performed better. These differences may reflect a key limitation: the pre-resection external surface represents the tissue state before resection, whereas the post-resection cavity reflects its state afterward. Aligning the specimen mesh across these two different deformation states can introduce inconsistencies.

Skin specimens typically experience relatively minor shrinkage, in contrast to the more pronounced mucosal shrinkage observed in oral specimens (see Fig.~\ref{fig:deformation_formulation}). This reduces the discrepancy between the pre-resection surface and the post-resection cavity. Additionally, the pre and post-resection-site surfaces are approximately parallel, so additional constraints from both surfaces may offer little benefit. These factors may explain why the proposed method performed comparably to~\cite{Yang2025Nonrigid} in skin specimens. In buccal specimens, the surfaces are also parallel, limiting the benefits of the additional constraint. Furthermore, buccal specimens are prone to significant mucosal shrinkage, which increases the mismatch between the pre-resection external surface and the post-resection cavity, thereby reducing the effectiveness of the additional constraint. For tongue specimens with irregular and non-parallel structures, relying solely on the post-resection cavity is insufficient. Although the external surface constraint is not fully realistic, it provides valuable thickness information that guides the registration process. Our method shows promising improvement over the previous method in the challenging tongue cases.

Another limitation of our experimental setup is that data collection poses challenges for thinner specimens. The RGB-D camera cannot accurately capture the thickness information between the pre- and post-resection surfaces for buccal and skin specimens. This limitation reduces the reliability of the thickness constraint in guiding the proposed deformable registration. 

Lastly, we only have access to a limited number of cadaver specimens, limiting the strength of our deformation results. Nonetheless, our results show a significant improvement in TRE between rigid and nonrigid registration methods for relocalizing the margins of surgical specimens. These limitations highlight the challenges of registering specimens with complex deformation behaviors and underscore the need for further research to refine data acquisition methods and improve deformation consistency in registration. Future work could consider incorporating data-driven methods~\cite{wu2020leveraging,wang2024libr+} to inform deformable registration.

\textbf{User Study}
The user study only has four cases, limiting the significance of the conclusions that can be drawn. Nonetheless, the magnitude and consistency of improvement highlight the potential of the system to reduce relocation error, and thus potentially cancer recurrence rates~\cite{Orosco2018PositiveCancers,Miller2024HowNeck}.  We also want to highlight the ability of our framework to easily integrate into the clinical workflow. Our pipeline, running on a regular laptop, takes 30 minutes. Only 10 minutes of direct access to the specimen and post-resection cavity are needed for 3D scanning, which is done with mobile scanners that are transportable with one person. The remaining process can occur remotely while pathology analyzes the specimen, minimizing workflow interruption. Incorporating machine learning deformation models may further reduce the processing time~\cite{wu2021learning}. 

\section{Conclusion}
We adapted the deformable registration approach to account for the thickness of three specimen types by incorporating pre-resection external surface information. The effectiveness of this method was supported by our user study results. The proposed framework integrates deformable registration with AR visual guidance to support accurate positive margin relocation, which can help reduce cancer recurrence rates. A pilot study demonstrated the proposed framework's effectiveness and potential for smooth integration into current clinical workflows.  


    

\begin{credits}
\subsubsection{\ackname}  This work was supported in part by the NIBIB of the NIH Grant R01EB027498, NCI of the NIH Grant 1K08CA293255, and Vanderbilt University Seeding Success Grant.

\subsubsection{\discintname}
This work has not been submitted for publication elsewhere. The authors have no competing interests to declare that are
relevant to this article.
\end{credits}

%
%
%
\bibliographystyle{splncs04}
\bibliography{references}

\end{document}